# A hybrid video quality metric for analyzing quality degradation due to frame drop

Manish K Thakur[1], Vikas Saxena[2] and J P Gupta[3]

[1] Department of CSE/IT, Jaypee Institute of Information Technology
Noida, Uttar Pradesh 201307, India
*mthakur.jiit@gmail.com*

[2] Department of CSE/IT, Jaypee Institute of Information Technology
Noida, Uttar Pradesh 201307, India
*vikas.saxena@jiit.ac.in*

[3] Sharda University
Greater Noida, Uttar Pradesh 201306, India
*jaip.gupta@gmail.com*

**Abstract**
In last decade, ever growing internet technologies provided platform to share the multimedia data among different communities. As the ultimate users are human subjects who are concerned about quality of visual information, it is often desired to have good resumed perceptual quality of videos, thus arises the need of quality assessment. This paper presents a full reference hybrid video quality metric which is capable to analyse the video quality for spatially or temporally (frame drop) or spatio-temporally distorted video sequences. Simulated results show that the metric efficiently analyses the quality degradation and more closer to the developed human visual system.

## 1. Introduction

For last many years, usages of visual information (video) have been significantly increased due to advancements in digital technologies. Apart from contents a user is also concerned about quality of visual information which might be degraded while processing over it, likes acquisition, transmission, compression, editing etc [1-4]. Therefore it is always required to maintain the original visual quality while processing over visual information.

Analysis of quality degradation is the first step to ensure visual quality of processed data. Based upon availability of reference video, analysis of video quality degradations can be done in full reference (FR) where reference video is available along with distorted video, reduced reference (RR) where little information about reference video is available and no reference (NR) where only distorted video is available [5-7].

Processing over video sequence might introduce following distortions: spatial distortion (SD), temporal distortion (TD), and spatio-temporal distortion (STD). Spatially distorted video sequences have distortions only in spatial domain, temporally distorted video have distortions in temporal domain, and spatio-temporally distorted video sequences have distortions in both domains [8].

Degradation in video quality is analyzed either using subjective or video objective quality assessment [9]. Subjective quality assessment is done by involving many human subjects who rate the visual quality. It is a lengthy and costly process [10]. The remedy is objective quality assessments which involve quality metrics to determine the video quality. These metrics are mainly designed upon the observations during subjective analysis [11-13] i.e. objective quality metrics are mainly designed upon the recommendations of human visual system (HVS).

For last many years numerous FR, RR, and NR objective quality metrics are proposed by researches which are classified as either mathematical metrics or HVS based metrics. Mathematical metrics analyses the video quality using some statistical functions. HVS based metrics are designed based formulated functions using observations of human subjects. Peak signal to noise ratio (PSNR) and Mean square error (MSE) are mathematical metrics whereas Structural similarity index (SSIM) and Video quality metric (VQM) are HVS based metrics [12-16].

Apart from quality assessment due to spatial distortion, it is also required to analyse the quality deterioration due to temporal distortion. A video sequence can be temporally distorted by dropping frames or swapping frames or inserting average frames into reference video sequences [17-19]. These temporal distortions are context sensitive; therefore there impact over perceived visual quality varies in different context. For example, considering a video sequence in which adjacent frames are identical, frame drop will not have much impact over perceived quality whereas a video sequence in which there is frequent change of scene in adjacent frames, frame drop will have significant impact over perceived visual quality.

It had been observed and presented by authors in next section that two of the popular metrics PSNR and SSIM are inefficient to analyse the quality degradation due to temporal distortion which raise the need of a quality metrics capable to analyse the quality degradation due to temporal distortion.

Apart from introduction, the paper is organized as follows: a subjective model for temporal distortion is presented in section 2, a hybrid video quality framework has been presented in section 3 followed by simulation and analysis in section 4.

## 2. A Subjective Model

As analysis of visual quality due to temporal distortion is context sensitive, therefore we developed and presented a human visual system [20] and obtained the mean opinion score (MOS) under different scenarios and cases mentioned in subsequent paragraphs.

*Scenario 1:* If reference videos are only spatially distorted.
*Scenario 2:* If reference videos are only temporally distorted (frame drop).
*Scenario 3:* If reference videos are distorted both temporally and spatially.

We defined and tuned following parameters: contiguous frame drop (cfd) and total dropped frames (tdf). Based upon these parameters we had identified following cases:

*Case 2.1:* If reference videos are distorted with low contiguous frame drop and total dropped frames as low. We called cfd of 1 to 5% and tdf of 1 to 10% as low.
*Case 2.2:* cfd is low, and tdf is high (we called tdf of 11 to 20% as high).
*Case 2.3:* cfd is high (we called cfd of 6 to 10% as high) and tdf is low.
*Case 2.4:* cfd is high, and tdf is high.

Considering reference video $V_R$ as $\{V_{R1}, V_{R2}, .. V_{Rm}\}$, distorted video $V_D$ as $\{V_{D1}, V_{D2}, .. V_{Dn}\}$, and one chunk of the contiguous dropped frames is from $V_{Rj}$ to $V_{Rj+c}$ where $c$ is the count of contiguously dropped frames in $V_R$ from index j. We had identified following possibilities which represents different context in a video sequence:

*Possibility 1:* If $V_{Rj} \approx V_{Rj+c}$ (i.e. adjacent frames in between dropped frames are almost similar), and $V_{Rj+1}$ to $V_{Rj+c-1}$ are also almost similar to $V_{Rj}$ or $V_{Rj+c}$.

*Possibility 2:* If $V_{Rj} \approx V_{Rj+c}$ and $V_{Rj+1}$ to $V_{Rj+c-1} \neq V_{Rj}$ or $V_{Rj+c}$, as $V_{Rj} \neq V_{Rj+1}$, there is change of scene from previous non dropped frame to first dropped frame. Similarly there will be change of scene from last dropped frame $V_{Rj+c-1}$ to next non dropped frame $V_{Rj+c}$.

*Possibility 3:* If $V_{Rj} \neq V_{Rj+c}$ (i.e. adjacent frames in between dropped frames are different) and $V_{Rj+1}$ to $V_{Rj+c-1}$ are almost similar to $V_{Rj}$.

*Possibility 4:* If $V_{Rj} \neq V_{Rj+c}$ and $V_{Rj+1}$ to $V_{Rj+c-1} \neq V_{Rj}$, as $V_{Rj} \neq V_{Rj+1}$, there is change of scene from previous non dropped frame to first dropped frame. Similarly, there will be change of scene from last dropped frame $V_{Rj+c-1}$ to next non dropped frame $V_{Rj+c}$.

The subjective assessment had been conducted for scenario 2 and scenario 3 over five different publicly available video sequences (figure 6) named as Bus, Coastguard, Galleon, Football, and Stefan. The obtained MOS under different scenarios (scenario 3 is divided into scenario 3a and scenario 3b depending upon percentage of introduced spatial distortion) and cases are presented in figure 1, figure 2, and figure 3. In scenario 3a spatial distortion is less whereas in scenario 3b, spatial distortion is high.

In subjective experiments, it has been observed that drop of bigger chunk (*i.e.* high cfd) described as case 2.3 and case 2.4 is more experienced by subjects as compared with smaller chunks (*i.e.* low cfd) described as case 2.1 and case 2.2. Further possibilities 1 and 3 are lesser experienced by subjects as compared with possibilities 2 and 4 where there is an abrupt change observed in non-dropped adjacent frames.

It had been noticed that when the cfd was low in possibilities 1 and 2, the missing clip was less experienced by subjects as compared with high cfd. If tdf is low then it is less experienced by subject, but it experienced the early completion of video if tdf was high.

The distortions in scenario 2 and scenario 3 (with low SD) have experienced almost similar observations (figure 2 and figure 3), suggesting that low SD does not make any significant impact over the perceived quality under the low and high TD. But the deterioration in quality in scenario 3

(high SD) has been experienced maximum. The obtained observations may be interpreted that high SD has significant impact over the perceived quality irrespective of high and low TD.

In continuation of proposed HVS for frame drop distortion, we also analyzed whether PSNR and SSIM are capable enough to compute the quality degradation in temporally distorted (frame drop) video sequences or not.

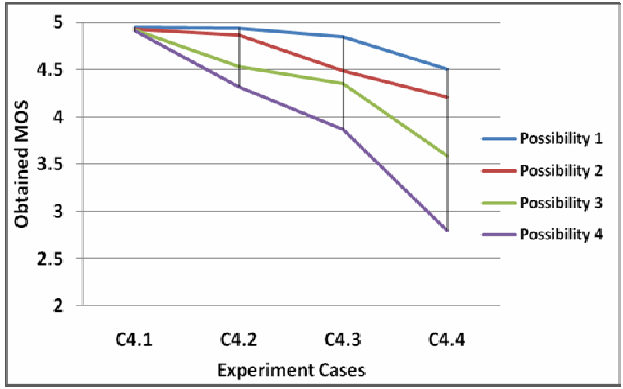

Fig. 1  MOS for different cases and possibilities under Scenario 2.

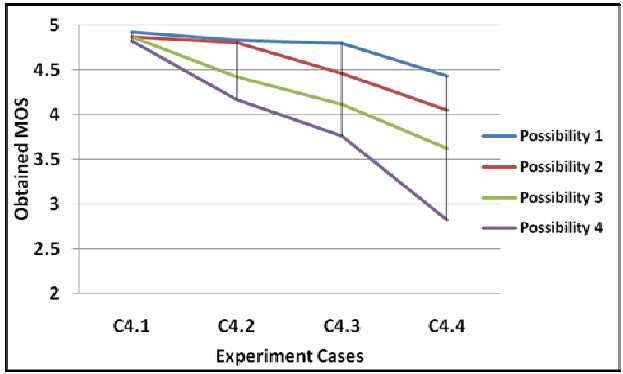

Fig. 2  MOS for different cases and possibilities under Scenario 3a.

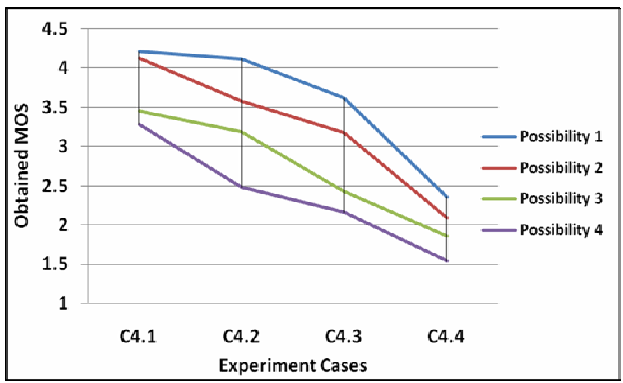

Fig. 3  MOS for different cases and possibilities under Scenario 3b.

It had been noted that due to unaligned video sequences, the quality degradation computation is unpredictable and thus arises a need to propose an objective quality metrics which efficiently computes the degraded quality due to temporal distortion (frame drop).

## 3. Dropped Frame Video Quality Metric

This section presents a full reference hybrid video quality metric, named as DFVQM (Dropped Frame Video Quality Metric) which by computing DFVQM index (DFVQMI) analyses the quality degradation in raw video sequences due to SD, TD, and STD.

As observed in section 2, perceptual distortion is more when; (a) there is frame drop in bigger chunk, *i.e.* cfd is too high, (b) total dropped frames (tdf) are high, and (c) there is significant SD i.e. human subjects are perceptually sensitive to the above mentioned parameters.

We mapped human (subjects) perceptions, as presented in previous section into our model and designed the framework over the identified sensitive parameters: high and low SD, high or low tdf, and high or low cfd.

As depicted in figure 4, the proposed framework firstly, identifies the lost frames indices, inserts average frames to those indices (such that videos are temporally aligned), computes the impact over quality due to SD and TD, and finally computes the quality index DFVQMI, as presented in Eq. (3). Subsequent paragraphs describe the steps involved while computing the quality degradation.

***Step 3.1:*** Input reference video sequence $V_R$ and distorted video sequence $V_D$ with m and n frames respectively, where, m ≥ n.

For example, consider $V_R$ with 12 frames (m=12) having indices as 1, 2, 3, 4, 5, 6, 7, 8, 9, 10, 11, 12 and $V_D$ with 6 frames (n=6) of $V_R$ having indices as 1, 2, 7, 8, 10, 12.

$V_R$ = | 1 | 2 | 3 | 4 | 5 | 6 | 7 | 8 | 9 | 10 | 11 | 12 |

$V_D$ = | 1 | 2 | 7 | 8 | 10 | 12 |

***Step 3.2:*** Analyse $V_R$ and $V_D$ for temporal distortion (drop of frames). If there is no temporal distortion then, copy $V_D$ into $V_C$, and go to *Step 3.4* otherwise continue.

***Step 3.3:*** Identify the indices of lost frames in $V_R$ with respect to $V_D$ (*i.e.* indices of frames in $V_R$ which are not available in $V_D$).

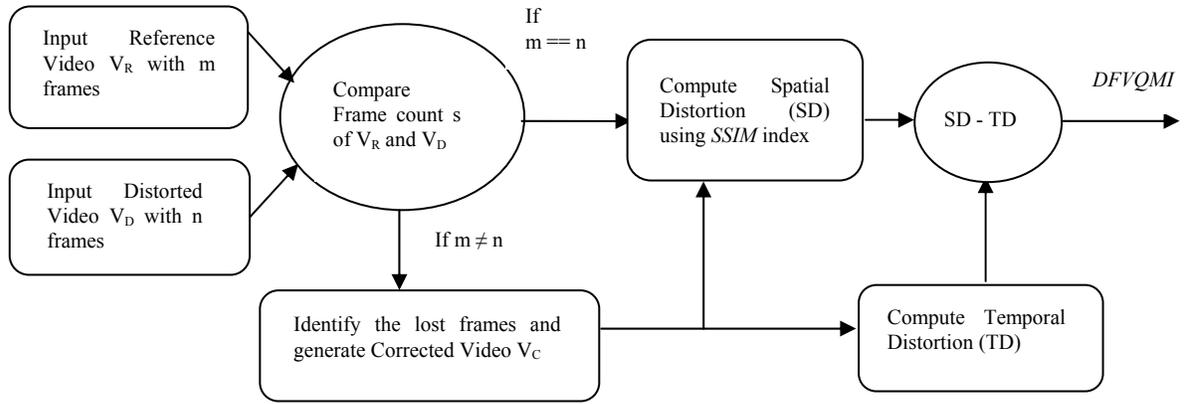

Fig. 4 The proposed framework for quality assessment.

Authors had presented an evolutionary algorithm to identify dropped frame indices in uncompressed video. Major steps of the algorithm are presented in figure 5. The algorithm performance had been analyzed using numerous test cases and it was noted that the scheme efficiently identifies dropped frame indices in maximum cases [21].

Store these dropped frames indices into an array Missing. Compute the contiguous frame drop count from identified lost frames indices and store these counts into an array Cd.

Therefore, in the preceding example, identified dropped frame indices stored into array Missing are 3, 4, 5, 6, 9, 11. Further, the array Cd will contain 4, 1, 1 corresponding to contiguously dropped frames 3 to 6, 9 to 9, and 11 to 11.

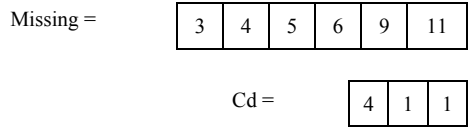

**Step 3.4:** Construct corrected video sequence $V_C$ by inserting average frames at identified dropped frames indices in $V_D$. If there is no dropped frames then $V_C = V_R$.

There can be different possibilities for generating the average frames and inserted into $V_C$ at identified indices. Case 3.4.1: at the identified indices, copy the last frame before a dropped frame of $V_D$ into $V_C$. Case 3.4.2: at the identified indices copy into $V_C$, the average of adjacent non-dropped frames of $V_D$. Case 3.4.3: copy into $V_C$, the contiguous average of adjacent frames of $V_C$ and $V_D$.

If there are large number of contiguous dropped frames then in case 3.4.1 all such frames will be replaced by single video frame which is the last non-dropped frame, hence human subject will not find motion in the video for longer time and video will look like a static video.

In similar way case 3.4.2 exhibits the contiguous dropped frames which are replaced by a single video frame; an average of two adjacent non-dropped video frames in between the contiguous dropped frames. This problem might not occur in case 3.4.3, but it will degrade the quality of the generated average frames if count of contiguous dropped frame is large.

For our framework, we have used case 3.4.1 as it is broadly used for inserting dropped packets in multimedia communication [22].

---

m= Count of frames in $V_R$; n = Count of frames in $V_D$
d = Count of dropped frames i.e. m - n
*DiffMat* [*n*] [*m+2*]   //stores PSNR, maximum PSNR, and index of maximum PSNR

**Step 1:** Input reference video $V_R$ and distorted video $V_D$

**Step 2:** Compute PSNR of $i^{th}$ frame of $V_D$ to $i^{th}$ to $(i+d+1)^{th}$ frame of $V_R$ for all i from 0 to n. Store these values into *DiffMat* up to $m^{th}$ field. Store maximum PSNR and its index of each row into $(m+1)^{th}$ and $(m+2)^{th}$ field of *DiffMat*.

**Step 3:** Compute the longest increasing sequence (LIS) in the index field (m+2) of *DiffMat*. Case 1: If length of LIS is n, then call the indices into LIS as non dropped frames and other indices as dropped frames and exit. Case 2: If length of LIS is less than n, go to step 3.

**Step 4:** Use mutation operator of genetic algorithm, create guided initial population, compute fitness score (FS) which is the average PSNR of all the indices included into initial population, generate next population, compute its FS, if new FS is better than previous, store the new population. Repeat the population generation up to $k^{th}$ step. After $k^{th}$ step, call the indices into population list as non dropped frames and other indices as dropped frame and exit.

---

Fig 5 Algorithm to compute dropped frame indices [21].

Therefore in the preceding example, the corrected video sequence $V_C$ is created with reference video indices as 1, 2, 2, 2, 2, 2, 7, 8, 8, 10, 10, 12 where frame number 2 is being copied to all the contiguously dropped frames ranging from index 3 to 6, frame number 8 is copied at the place of dropped frame index 9, and frame number 10 is copied at the place of dropped frame index 11. In the preceding example all the indices in $V_R$, $V_D$, Missing, and $V_C$ are the indices of reference video $V_R$.

$V_C$ = | 1 | 2 | 2 | 2 | 2 | 2 | 7 | 8 | 8 | 10 | 10 | 12 |

**Step 3.5:** Compute the spatial distortion SD by averaging SSIM of all non-dropped frames in $V_C$ and $V_R$ (Eq. (1)). We used SSIM over PSNR for computing SD because SSIM performance is better than PSNR [9, 20]. Therefore, if $V_R.Length$ (*i.e.* m) equals to $V_D.Length$ (*i.e.* n) then, SD is the average SSIM of all frames of $V_R$ and $V_D$, otherwise, it will be average SSIM of non dropped frames.

$$SD = \frac{1}{n}\sum_{i=1}^{m} SSIM(V_{R_i}, V_{C_i}) \quad V_{Ci} \neq \text{Missing Frame} \quad (1)$$

**Step 3.6:** Compute temporal distortion TD using Eq. (2)

$$TD = \sum_{i=1}^{cd.Length} \frac{cd_i}{m} \times Avg(C_1 + C_2) \quad (2)$$

where,

$$C_1 = (1 - SSIM(V_{C_{j-1}}, V_{C_j + cd_i}))$$

$$C_2 = (1 - \frac{1}{cd_i} \times \sum_{t=j+1}^{j-1+cd_i} SSIM(V_{R_t}, V_{C_t}))$$

j = Missing$_k$, initially for k =0 and incremented by k + $cd_i$

**Step 3.7:** Compute the overall quality degradation using our proposed index DFVQMI as given in Eq. (3).

$$DFVQMI = SD - TD \quad (3)$$

## 4. Simulation and Analysis

In previous section we presented the quality index DFVQMI which analyses the quality degradation by subtracting TD from SD. If there is no frame drop then the computed quality index DFVQMI will be simply the frame by frame difference of $V_R$ and $V_C$ using SSIM.

If there are drop of frames then the TD will be computed according to Eq. (2) which contains following literals: $cd_i/m$, $C_1$, and $C_2$.

The literal $cd_i/m$ gives the temporal deterioration in a reference video of length m due to drop of one chunk of dropped frames. $C_1$ and $C_2$ control temporal distortion to behave like four possibilities of each case in scenario 2 and scenario 3 in section 2. Subsequent paragraph emphasize the impact over temporal distortion due to $C_1$ and $C_2$:

If $V_{Cj-1} \approx V_{Cj+cdi}$ then $C_1$ will be approaching to 0 resulting in decrement of TD whereas if $V_{Cj-1} \neq V_{Cj+cdi}$ then $C_1$ will be approaching to 1 resulting in increment of TD. If $V_{Cj+1}$ to $V_{Cj+cdi-1} \neq V_{Rj}$ then $C_2$ will be approaching to 1 resulting in increment of TD whereas if $V_{Cj+1}$ to $V_{Cj+cdi-1} \approx V_{Rj}$ then $C_2 \approx 0$, resulting in decrement of TD

Averaging of $C_1$ and $C_2$ controls the increment or decrement of TD in all the four possibilities and $cd_i/m$ controls its rise or fall with respect to cfd and tfd.

In worst case, $C_1$ is 1, $C_2$ is 1 and $cd_i/m$ is 1, thus TD is 1, and if SD is 0, DFVQMI is -1, whereas in best case, TD is 0 and SD is 1, thus DFVQMI is 1 i.e. the index range of DFVQMI is from -1 to +1.

We simulated the proposed framework and analyzed its performance using five publicly available video sequences (figure 6) which were also used in subjective assessment.

To analyse the performance of DFVQMI for scenario 2, we distorted the video sequences by dropping video frames (starting frames, or middle frames, or last frames) such that cfd ranges from 1% - 10% (*i.e.* 2 to 25 frames of each video) and total dropped frames ranges from 1% - 20% *(i.e.* 2 to 50 frames).

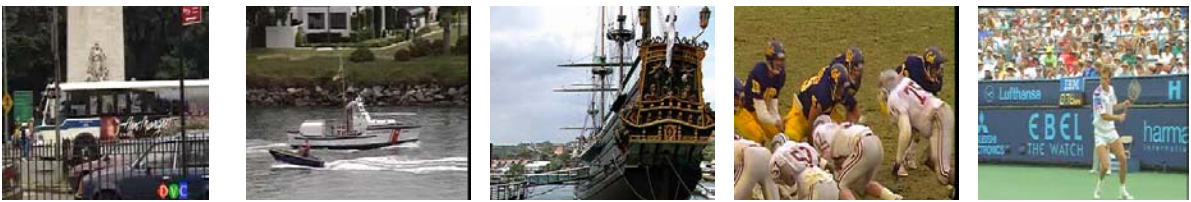

Fig. 6 Video sequences (http://media.xiph.org/video/derf/) used as test videos.

Further to analyse the performance of DFVQMI for scenario 3, we intentionally introduced SD at specific positions (at LSB for low SD, and $4^{th}$ LSB for high SD) using LSB watermarking [23] in the same dataset used in scenario 2 having TD.

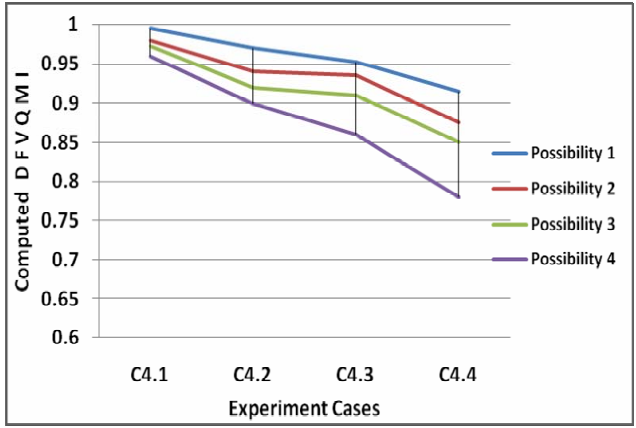

Fig. 7 DFVQMI for different cases and possibilities under Scenario 2.

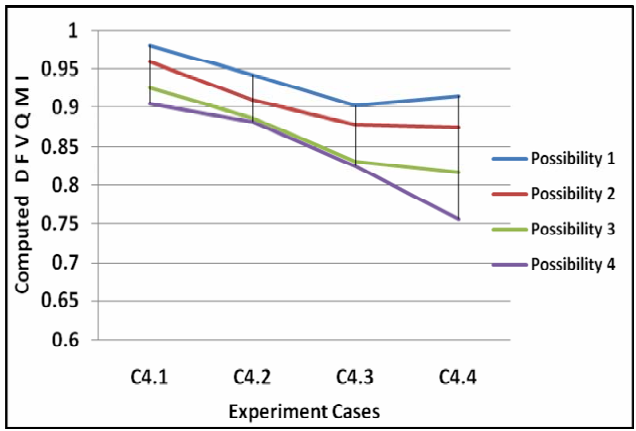

Fig. 8 DFVQMI for different cases and possibilities under Scenario 3a.

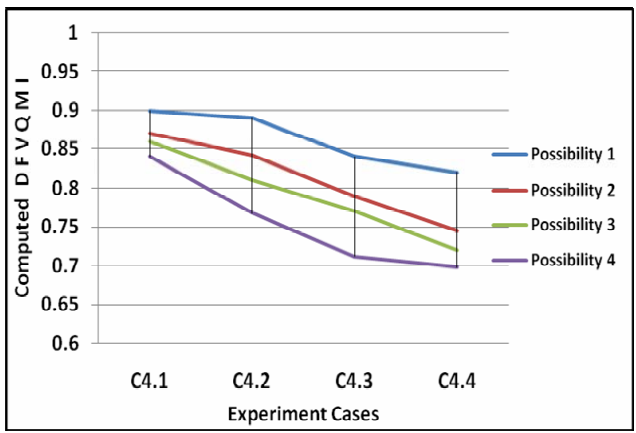

Fig. 9 DFVQMI for different cases and possibilities under Scenario 3b.

In total, we used 5 reference videos and 80 distorted videos (5 reference video × 4 cases × 4 possibilities) for scenario 2; whereas we used 5 reference videos and 160 distorted videos (80 distorted videos of scenario 2 × 2 types of SD including LSB and $4^{th}$ LSB Watermarking) for scenario 3.

After conducting numerous experiments the analyzed quality degradation by DFVQMI has been presented in figure 7, figure 8, and figure 9 under scenario 2, scenario 3a, and scenario 3b respectively.

It has been observed that the computed degradation by DFVQMI is closer to the obtained subjective scores in section 2. But obvious reason is the aligned distorted video sequence before getting it compared with reference video. Further, impact of temporal distortion under different context has been computed and subtracted from overall degradation due to spatial distortion brings the overall quality degradation more closer to the obtained MOS.

## 4. Conclusions

In this paper we explained the need of objective video quality metrics which will be capable to analyse the quality degradation due to temporal and spatio-temporal distortion (frame drop) and presented a framework which computes the dropped frame video quality metric index (DFVQMI) which is ranged between -1 to +1. Performance of the proposed framework has been tested for different contexts and it has been observed that the proposed framework is capable to analyse the quality degradation due to spatial, temporal, and spatio-temporal distortions. It has been also noted that computed degraded quality using framework index DFVQMI is closer to the obtained MOS.

**Manish K Thakur** is currently working as Senior Lecturer at JIIT Noida, India. He did his M Tech from BIT Mesra in 2004 and currently pursuing his Ph D from JIIT Noida. He is author of more than 6 research papers published in International journals and conferences. His research interests are in the field of video processing, data structures and algorithms.

**Vikas Saxena** did his Ph D in 2009 and currently working as Assistant Professor at JIIT Noida, India. He has more than 20 publications in International journals and conferences. His expertise is in the field of Image processing, computer graphics and computer vision.

**Prof. J P Gupta** is Vice Chancellor of Sharda University, Greater Noida, He is an academician having more than 35 years experience including IIT Roorkee, JIIT Noida, and Galgotia University, Greater Noida. He is author of more than 60 research papers.